# Observation of a Reconstructed Chern Insulator in Twisted Bilayer MoTe$_2$

Min Wu[1,2*], Lingxiao Li[1*†], Yunze Ouyang[3*], Yifan Jiang[4], Wenxuan Qiu[3], Zaizhe Zhang[1], Zihao Huo[1], Qiu Yang[1], Ming Tian[5], Neng Wan[5,8], Kenji Watanabe[6], Takashi Taniguchi[7], Shiming Lei[4†], Fengcheng Wu[3†], Xiaobo Lu[1‡]

[1]International Center for Quantum Materials, School of Physics, Peking University, Beijing 100871, China
[2]Lab of Low Dimensional Magnetism and Spintronic Devices, School of Physics, Hefei University of Technology, Hefei, Anhui 230009, China
[3]School of Physics and Technology, Wuhan University, Wuhan 430072, China
[4]Department of Physics, Hong Kong University of Science and Technology, Clear Water Bay, Hong Kong SAR, China
[5]Key Laboratory of MEMS of Ministry of Education, College of Integrated Circuits Southeast University, Nanjing 210096, China
[6]Research Center for Electronic and Optical Materials, National Institute of Material Sciences, 1-1 Namiki, Tsukuba 305-0044, Japan
[7]Research Center for Materials Nanoarchitectonics, National Institute of Material Sciences, 1-1 Namiki, Tsukuba 305-0044, Japan
[8]State Key Laboratory of Surface Physics, Key Laboratory of Micro and Nano Photonic Structures (MOE), and Department of Physics, Fudan University, Shanghai, 200433, China

*These authors contributed equally.
Email: lilingxiaodh3111@pku.edu.cn, phslei@ust.hk, wufcheng@whu.edu.cn, xiaobolu@pku.edu.cn

**Twisted bilayer MoTe$_2$ (tMoTe$_2$) is a prototypical moiré material in which long-wavelength superlattices amplify electron correlations, enabling a wealth of emergent quantum phases. To date, experimental efforts have focused primarily on small twist angles (typically < 4° ), whereas the larger-angle regime—where moiré bands become more dispersive and correlations are reduced—has remained largely unexplored. Here we chart the topological phase space of tMoTe$_2$ at a relatively large twist angle of approximately 4.54°, accessing a moderately correlated regime with enhanced bandwidth. In contrast to small-angle devices that predominantly host fractional quantum anomalous Hall or spin Hall responses, we uncover multiple Chern-insulating states with |$C$| = 1 at moiré fillings $v$ = −1, −0.53 and −1/2. Strikingly, at $v$ = −2/3 a magnetic field induces a fractional Chern insulator accompanied by an insulator–metal transition. Our results broaden the topological phase diagram of tMoTe$_2$ and establish large-angle moiré superlattices as a versatile platform for engineering robust topological states beyond the strong-correlation limit.**

**Main**

R-stacked twisted bilayer MoTe$_2$ (tMoTe$_2$) has emerged as a model moiré platform for exploring correlated and topologically nontrivial phases of matter[1-4]. At twist angles below 4°, a diverse set of quantum phases have been reported, including integer and fractional quantum anomalous Hall (IQAH, FQAH) states[5-14], zero-field gapless composite Fermi liquid[7,15], fractional topological insulator[16,17], and unconventional superconductivity proximate to the FQAH state[14], etc. These results point to an exceptionally rich correlated topological landscape in the strongly correlated regime of relatively narrow moiré bands. Because both the electron interaction strength and quantum geometry of the low-energy moiré bands are sensitive to the variations in the twist angle[1-4,18-25], a central question is how this landscape evolves when the moiré bands become more dispersive at larger twist angles, corresponding to a moderately correlated regime, which has remained largely unexplored experimentally. In this work, we address this regime by investigating the electronic properties of tMoTe$_2$ at a twist angle of approximately 4.54°. Transport measurements reveal several emergent phenomena, including electric-field-enhanced IQAH, previously unexplored QAH states at commensurate and incommensurate fillings, and magnetic-field-induced insulator-metal transition and re-entrant fractional Chern insulating state. Our results demonstrate a distinct, field-tunable regime of correlated topology beyond the strongly correlated limit.

**Phase Diagram**

We perform electrical transport measurements on dual-gated tMoTe$_2$ devices (schematically illustrated in Fig. 1a, See Method for details of the device fabrication). Figures 1c and 1d show the longitudinal resistance $R_{xx}$ and the Hall resistance $R_{xy}$ of device D1 as a function of filling $v$ and electric-field $D/\varepsilon_0$ at the base temperature $T = 10$ mK (unless otherwise specified, the data presented in this work are mainly collected on device D1), respectively. $R_{xx}$ and $R_{xy}$ are symmetrized and antisymmetrized under a small applied out-of-plane magnetic field $B = \pm 10$ mT to suppress accidental fluctuations. The moiré density is evaluated to be $n_M \approx 5.85 \times 10^{12}$ cm$^{-2}$ (one hole per moiré unit cell, $v = -1$), corresponding to a twist angle of approximately 4.54°, larger than those studied in previous works[5-17]. In this case, the phase diagram of the first moiré valence band can still be partitioned into two distinct regions: layer-hybridized region and layer-polarized region, which are separated by the tilted stripe regions (Fig. 1c). Within the layer-hybridized region, substantial $R_{xy}$ approaching $h/e^2$ accompanied by vanishing $R_{xx}$ (< 50 Ω) emerges at $v = -1$ (Figs. 1c, 1d, as indicated by black dashed lines), a hallmark of IQAH state, where $h$ and $e$ are the Planck constant and elementary charge, respectively. These observations are consistent with earlier reports on tMoTe$_2$ with twist angles below 4°[7,8,11-14]. Notably, the FQAH insulators at $v = -2/3$ and $-3/5$[7,8,13,14] are absent, likely due to the dispersive band structure and non-uniform Berry curvature distribution at larger twist angles[2,4,20-25]. Instead, at commensurate filling $v = -1/2$ and incommensurate filling $v = -0.53$ (as indicated by grey dashed lines in Figs. 1c, 1d), we observe large $R_{xy}$ accompanied by $R_{xx}$ minima (Extended data Fig. 6), indicative of topological non-trivial states that spontaneously break time-reversal and translational symmetries. In the layer-polarized region, several topologically trivial correlated insulating states are identified at $v = -1, -2/3, -1/2, -1/3$ and $-1/4$ (Fig. 1c and Extended Data Fig. 3).

**Collapsed IQAH state near zero electric field at $v = -1$**

We first examine the characterization of IQAH state at $v = -1$. Figure 1e shows $R_{xy}$ and $R_{xx}$ as a function of the magnetic field around $v = -1$ at $D/\varepsilon_0 = 0.15$ V/nm and $T = 10$ mK. Both the maximum of $R_{xy}$ and minimum of $R_{xx}$ vary linearly with the applied magnetic field down to zero, yielding a Chern number $|C| = 1$ according to the Streda formula $C = \Phi_0 \partial n/\partial B$ (indicated by the white dashed lines), where $\Phi_0$ is the magnetic flux quantum. The zero-field Hall resistance remains quantized up to 2 K, accompanied by the small longitudinal resistance and pronounced magnetic hysteresis loops, indicative of robust ferromagnetic order, as depicted in Fig. 1f. At higher temperatures, the Hall resistance $R_{xy}$ gradually deviates from the quantization, and the corresponding $R_{xx}$ begins to surge sharply. Based on the temperature dependence of $R_{xy}$ and $R_{xx}$ (Fig. 1f), the Curie temperature is estimated to be about 6 K at $v = -1$, smaller than that obtained in earlier works[7,8,11-14], likely originating from the weaker interaction strength at the larger twist angle.

Notably, the stability of the IQAH state exhibits a counterintuitive dependence on the out-of-plane electric field. As the electric field approaches zero, the range of filling factors ($v$) over which the IQAH state is observed narrows (see Figs. 1c and 1d). A similar electric-field-dependent suppression of the topological state is also observed in device D2, although without clear Hall quantization (Extended Data Fig. 1b). To further elucidate this behavior, we plot the electric-field-dependent IQAH state across various filling factors in Extended Data Fig. 4. Figure 2a plots $R_{xx}$ and $R_{xy}$ versus $D/\varepsilon_0$ at $v = -1$. At $T = 10$ mK and zero electric field, $R_{xy}$ exhibits a slight deviation from perfect quantization, accompanied by a marginal increase in $R_{xx}$ to $\sim 200$ Ω. Nevertheless, the values remain well within the IQAH effect framework compared with previous literature benchmarks[7,8]. By contrast, upon raising the temperature to 1.2 K, deviations from quantized Hall resistance become more pronounced, indicating the reduced robustness of IQAH state at zero electric-filed limit.

The collapse of IQAH state reflects a reduction of incompressible gap with $D/\varepsilon_0$. To quantify this effect, we examine the temperature dependence of $R_{xx}$ and $R_{xy}$ at selected electric fields, as shown in Fig. 2b. From the onset of IQAH effect breakdown, we extract the effective incompressible gap as a function of electric field in Fig. 2c (See Extended Data Fig. 5 for details). Although the magnitude of gap determined from $R_{xx}$ and $R_{xy}$ is different, both reveal the same nonmonotonic evolution with electric field: the gap increases from 1.8 K at zero electric field to a maximum of 11.3 K at 0.154 V/nm and subsequently decreases to 5.2 K at 0.253 V/nm for $R_{xx}$ case. Notably, as shown in Fig. 2b, $R_{xy}$ decreases monotonically with increasing temperature, evolving from a quantized value towards zero, whereas $R_{xx}$ initially displays nonlinear rise, peaks near temperature where $|dR_{xy}/dT|$ is maximal, and subsequently undergoes a resistive state-to-metal transition as evidenced by a sign change in $dR_{xx}/dT$. The nonmonotonic temperature dependent $R_{xx}$ is phenomenologically reminiscent of the Kondo-like transport behavior[26-29], but present data cannot verify the Kondo mechanism. Such a clear Chern insulator-resistive state-metal transition has not been previously reported in tMoTe$_2$, and calls for further theoretical investigations to reveal the microscopic origin.

Both the collapse of the IQAH state stability window and the nonmonotonic gap variation sharply contrast with the behaviors observed in tMoTe$_2$ devices with twist angles below 4°, where the $v = -1$ IQAH state remains robust near zero electric fields[7,8,11-14]. In contrast to the smaller twist angle devices in the strongly correlated regime, the moiré bands at larger twist angles are more dispersive, leading to a more pronounced variation in the density of states (DOS). As shown by the calculated single-particle DOS at a twist angle of 4.54° (Fig. 1b), as the interlayer potential difference $|U_z|$ decreases, the van Hove singularity (vHS), where the DOS diverges, shifts continuously from $v < -1$ to $v > -1$. Therefore, the vHS crosses the $v = -1$ at finite electric fields, coincident with the enhancement of the Curie temperature, as Stoner ferromagnetism is strengthened by the increase in the DOS. Similar enhancements of correlated insulators through field-tunable vHS have been observed in twisted bilayer WSe$_2$, although the insulators in that system are topologically trivial[30,31].

**IQAH states at $v = -1/2$ and $v = -0.53$**

We next investigate the previously unexplored topological states at fillings $v = -1/2$ and $v = -0.53$, which manifest as pronounced maxima in $R_{xy}$ accompanied by minima in $R_{xx}$ (Figs. 1c, 1d and Extended data Fig. 6). To further explore the topologically nontrivial properties of these two states, $R_{xx}$ and $R_{xy}$ versus $v$ and $B$ at $D/\varepsilon_0 = 0$ and $T = 10$ mK are presented in Figs. 3a and 3b, respectively. In addition to the $v = -1$ state, which exhibits a slope corresponding to a Chern number $|C| = 1$ (white solid lines), the $v = -1/2$ state displays a similar linear shift in the $v$–$B$ maps down to zero magnetic field, yielding the same Chern number $|C| = 1$ according to the Streda formula (white dashed lines). This observation is distinct from the physics reported in smaller twist angle systems (< 4°), where the $v = -1/2$ state appears to be a gapless composite Fermi liquid behavior[7,15]. The emergence of an integer Chern number at a commensurate fractional filling points to the interaction-driven quantum anomalous Hall crystal (QAHC), also known as topological charge density wave, stabilized by spontaneous breaking of translational symmetry[23,32-35]. To theoretically elucidate this state, we perform exact diagonalization (ED) calculation at filling $v = -1/2$, assuming spontaneous full spin polarization. We find four quasi-degenerate many-body ground states separated by a finite gap from excited states (Fig. 3g), with momentum indices consistent with a charge density wave state (Fig. 3h). Under twisted boundary condition, we evaluate the many-body Chern number and find that each of the quasi-degenerate ground states carries $|C| = 1$, in agreement with the observed IQAH transport signatures. The real-space hole density (Fig. 3i), calculated for a state that is a linear superposition of the four quasi-degenerate many-body ground states pinned by an attractive impurity potential, exhibits a pronounced 2 × 2 charge modulation. The maxima are located on one out of four MM sites (i.e., the high-symmetry sites where the metal atoms are vertically aligned) within the moiré superlattice, providing direct evidence for the crystalline order. Thus, the IQAH state observed at $v = -1/2$ is compatible with the QAHC scenario, consistent with the theoretical proposal for tMoTe$_2$[23]. This topological phase has also been experimentally observed in rhomobohedral pentalayer graphene and twisted bilayer–trilayer graphene[36,37], but has not, to our knowledge, been revealed previously in semiconductor moiré superlattices[7,8,11-14].

While $R_{xy}$ is not fully quantized to $h/e^2$ with a finite $R_{xx}$ (~ 2 kΩ, Extended Data Fig. 6) under zero magnetic field at $v = -1/2$, it acquires ~ 97% quantization under a small magnetic field of $|B| = 10$ mT, and exhibits evident hysteresis loop, as shown in Fig. 3c. This suggests incomplete suppression of dissipative channels of our device, which likely arises from limited bulk mobility, local strain or twist angle inhomogeneity rather than the absence of underlying topological order. The QAHC state remains stable up to ~ 250 mK, above which the anomalous Hall response gradually deviates from quantization (Figs. 3c and 3d), while the ferromagnetic order persists to a Curie temperature of $T_c \approx 1.5$ K.

Upon slight hole doping beyond $v = -1/2$, an additional topological phase occurs at the incommensurate filling $v = -0.53$ (Fig. 3e), and this phase exhibits an anomalous magnetic field dependence (Figs. 3a, 4a, 4b, 4e, and Extended Data Fig. 6). In contrast to $v = -1$ and $v = -1/2$, neither $R_{xx}$ minimum nor $R_{xy}$ maximum displays a well-defined linear dispersion in $v$–$B$ map at $v = -0.53$. A linear fit of the $R_{xx}$ minimum trajectory at small magnetic fields, when interpreted via the Streda formula, would indicate an effective Chern number substantially larger than $|C| = 1$ (Extended Data Fig. 6). Despite this unconventional field dependence, a quantized $R_{xy}$ (~ 0.99 $h/e^2$) with $R_{xx}$ ~ 1 kΩ under $|B| = 10$ mT and $T = 10$ mK is observed, as shown in Fig. 3e and Extended Data Fig. 6. Such behavior at an incommensurate filling is reminiscent of the re-entrant IQAH effect at $v = -0.63$ in 3.83° tMoTe$_2$[14], the extended QAH effect in rhombohedral multilayer graphene[38], and the re-entrant integer quantum Hall effects in high-mobility two-dimensional electron gases at high magnetic fields[39,40]. This incommensurate IQAH state persists up to ~ 150 mK, whereas the associated ferromagnetism disappears near $T = 1$ K. Given that the QAHC states are theoretically explored predominantly at commensurate fractional fillings[23,32-35], the emergence of QAH state with anomalous magnetic field dependence at incommensurate filling suggests a different microscopic mechanism (beyond the scope of present work) that warrants further theoretical and experimental explorations.

**Magnetic-Field-Driven Phase Transition at $v = -2/3$**
Figures 4a and 4b show the $v$–$D/\varepsilon_0$ maps of $R_{xx}$ near $v = -2/3$ at different magnetic fields and base temperature $T = 10$ mK (the corresponding $R_{xy}$ is shown in Extended Data Fig. 7). The resistance peaks emerging in the range 0.114 V/nm < $D/\varepsilon_0$ < 0.167 V/nm (Fig. 1c) of the layer-polarized regime are strongly suppressed (Fig. 4a) and completely annihilated (Fig. 4b) at $|B|$ = 0.25 T and 1 T, respectively. This magnetic field driven evolution is further illustrated in Fig. 4c, which plots $R_{xx}$ against $v$ and $B$. The resistance peak diminishes continuously as the magnetic field increases and disappears beyond ~ 320 mT. A similar phenomenon is observed upon increasing temperature, indicative of a melting behavior (Extended Data Fig. 7). From the temperature dependence of longitudinal resistance, we extract the energy gap by fitting the data to the Arrhenius equation at various magnetic fields (Fig. 4d), which displays a significant suppression with increasing the magnetic field. These observations suggest that the correlated insulating state at $v = -2/3$ sharply contrasts to the conventional expectation for charge density wave order, which is typically stabilized by magnetic fields[41,42]. Instead, this anomalous magnetic response may arise from local magnetic scattering, potentially due to spin fluctuations[12,13] or the Kondo-like physics[26-29].

At higher magnetic fields, a new phase emerges near $|D/\varepsilon_0|$ = 0.1 V/nm within the layer-hybridized regime, characterized by a small $R_{xx}$ accompanied by a large $R_{xy}$, as shown in Figs. 4e and 4f, indicative of a magnetic field induced topologically nontrivial state. A linear dispersion originating from $v = -2/3$ is identified according to the Streda formula, giving rise to a Chern number $|C| = 2/3$, as denoted by the black dashed line in Fig. 4g (the corresponding $R_{xx}$ is plotted in Extended Data Fig. 7). This topological state bears close resemblance to the re-entrant IQAH effect observed in 3.83° tMoTe$_2$[14]. However, in comparison to the zero-field quantized counterpart at smaller twist angle, the present re-entrant state remains imperfectly quantized even under the magnetic field up to 4 T (Fig. 4h). The Chern number of this re-entrant topological state is $|C| = 2/3$ according to the Streda formula, rather than $|C| = 1$ (Extended Data Fig. 7), indicates a putative fractional Chern insulating state, reminiscent of that observed in graphene moiré systems[43].

**Discussion and Conclusion**
Our results uncover distinct correlated topological phase diagram in 4.54° tMoTe$_2$, deviating from the established phenomenology at smaller twist angles[5-17]. The nonmonotonic electric field dependence of $v = -1$ IQAH gap underscores the importance of vHS in strengthening topological states in the moderately correlated moiré bands, suggesting that engineering the alignment between the chemical potential and divergent DOS in topologically nontrivial bands provides a promising strategy for enhancing topological phases. The emergence of unconventional topological states at $v = -1/2$ and $v = -0.53$, as well as the magnetic field assisted re-entrant topological state at $v = -2/3$, demonstrates that the topological physics is reshaped at moderately correlated regime compared to the strongly correlated regime.

In conclusion, we have mapped the electronic phase diagram of tMoTe$_2$ at twist angle of approximately 4.54°, unveiling a rich landscape of interaction-driven topology, in which commensurate and incommensurate fillings host different correlated topological states. These findings underscore the importance of twist angle and band structure engineering in realizing robust and diverse quantum phases in semiconductor moiré superlattices. Future experiments combining local scanning probes with transport measurements will be essential to directly visualize the translational symmetry breaking of the QAHC state and verify magnetic field assisted re-entrant fractional Chern insulator.

**Note**: A recent study of twisted monolayer-bilayer MoTe$_2$ at a twist angle of 3.7° reported similar phenomena, including a QAHC at $v = -1/2$ and an incommensurate IQAH state near $v = -0.53$[44].

**Methods**
**Device Fabrication**
Hexagonal boron nitride (hBN), graphite and 2H-MoTe$_2$ (HQ) flakes were exfoliated from bulk crystals onto 285 nm SiO$_2$/Si substrates with the thickness determined by optical contrast and atomic force microscopy (AFM). The tMoTe$_2$ devices with local contact gate were fabricated using dry-transfer and "cut and stack" method[7]. The bottom gate was prepared by picking up a hBN flake (thickness 10−15 nm) and a few-layer graphite flake using polycarbonate/polydimethylsiloxane stamp, then released onto the SiO$_2$/Si substrate at a temperature of 180 °C. Contact electrodes Pt (~ 8 nm) were patterned into Hall-bar configuration by standard electron beam lithography (EBL) and e-beam evaporation, followed by annealing at 350 °C for 3 hours and cleaning under AFM contact mode to remove the residue. To prevent oxidation or degradation, the tMoTe$_2$ structure was fabricated in a glovebox with O$_2$ and H$_2$O concentrations below 0.1 ppm. A MoTe$_2$ monolayer was cut into two parts by AFM tip or laser and sequentially picked up by a thin hBN (~ 5 nm) with a twist angle of 4.54°. The finished stack was released onto Pt electrodes at 180 °C. A local Pt contact gate (~ 8 nm) was deposited on the thin hBN above the contact electrodes to obtain a heavy doping and reduce the contact resistance, which was cleaned by AFM again. The top gate made of graphite/hBN (~ 10 nm) was fabricated to complete the transfer. Next, Cr/Au (5/60 nm) contacts were evaporated to connect Pt and the prepatterned outer pads. Finally, additional EBL and reactive ion etching were performed to eliminate the potential transport signals from MoTe$_2$ monolayer.

**Transport Measurement**
Transport measurements were conducted in a dilution refrigerator (Oxford Triton 200) with a base phonon temperature of about 10 mK. Standard low frequency (9.777 Hz) lock-in techniques with voltage preamplifier (input resistance of 100 MΩ) were used to measure the device resistance. The excitation current is 2nA to minimize sample heating.

**Twist Angle Calibration**
The resistance map measured as a function of top ($V_{tg}$) and bottom gate ($V_{bg}$) voltages (Extended Data Fig. 3a) was transformed into the $n-D/\varepsilon_0$ parameter space (Extended Data Fig. 3b). The carrier density $n$ and perpendicular electric field $D/\varepsilon_0$ are defined by top and bottom gate voltages according to $n = (C_{tg}V_{tg} + C_{bg}V_{bg})/e - n_{offset}$ and $D/\varepsilon_0 =$

$(C_{tg}V_{tg} - C_{bg}V_{bg})/2\varepsilon_0 - D_{offset}$, where $\varepsilon_0$, $n_{offset}$ and $D_{offset}$ are vacuum permittivity, intrinsic doping and built-in electric field, respectively. $C_g = \varepsilon\varepsilon_0/d_g$ is the gate capacitance per unit area with $\varepsilon \approx 3$ and $d_g$ the relative permittivity and thickness of hBN. $D_{offset} \approx 10$ mV/nm is estimated from the dual-gate map (Extended Data Fig. 3a). By finely tuning the ratio between top and bottom capacitances that align the resistance peaks parallel to the electric field axis, three most pronounced resistance peaks (indicated by black dashed lines) are clearly identified at the filling factors $v = -1, -2/3$ and $-1/2$. Thus, the moiré density is extracted to be $n_M \approx 5.85 \times 10^{12}$ cm$^{-2}$, corresponding a twist angle of $\theta = a\sqrt{\frac{\sqrt{3}}{2}n_M} = 4.54°$, where $a = 0.352$ nm is the lattice constant for monolayer MoTe$_2$[45].

**Single-Particle Model**

The moiré Hamiltonian for the valence bands in the $\pm K$ valleys of tMoTe$_2$ was constructed in Ref[1].

$$\widehat{\mathcal{H}}_0^\tau = \begin{pmatrix} \frac{-\hbar^2(\widehat{k} - \tau\kappa_+)^2}{2m^*} + \Delta_+(r) & \Delta_{T,\tau}(r) \\ \Delta_{T,\tau}^\dagger(r) & \frac{-\hbar^2(\widehat{k} - \tau\kappa_-)^2}{2m^*} + \Delta_-(r) \end{pmatrix}$$

$$\Delta_\pm(r) = 2V \sum_{j=1,3,5} \cos(g_j \cdot r \pm \psi) \pm U_z,$$

$$\Delta_{T,\tau}(r) = w(1 + e^{-i\tau g_2 \cdot r} + e^{-i\tau g_3 \cdot r}),$$

(1)

where $\widehat{\mathcal{H}}_0^\tau$ is the 2×2 Hamiltonian expressed in the layer-pseudospin space, and $\tau = \pm$ is the index of $\pm K$ valleys locked to spin ↑ and ↓, respectively. The layer-dependent potentials $\Delta_\pm(r)$ have the amplitude parameter $V$ and phase parameters $\pm\psi$, while the interlayer tunneling $\Delta_{T,\tau}(r)$ has a strength parameter $w$. $r$ and $\widehat{k} = -i\partial_r$ are the position and momentum operators, respectively. $\pm U_z$ is the layer-dependent potential induced by an external displacement field. $m^*$ is the effective mass. $\kappa_\pm = [4\pi/(3a_M)](-\sqrt{3}/2, \mp 1/2)$ are located at the moiré Brillouin zone corners, and the moiré reciprocal lattice vectors $g_j = [4\pi/(\sqrt{3}a_M)]\{\cos[(j-1)\pi/3], \sin[(j-1)\pi/3]\}$ with $j = 1, \cdots, 6$. Here, $a_M = a/\theta$ is the moiré period and $a$ is the monolayer lattice constant. We take the model parameters $a = 3.52$ Å, $m^* = 0.62\, m_e$, $V = 11.2$ meV, $\psi = 91.0°$, $w = 11.3$ meV from Ref[4], and $\theta = 4.54°$. Here $m_e$ is the bare electron mass.

By diagonalizing Eq. (1) in the plane wave basis, we obtain the single-particle moiré band with energy $\varepsilon_{nk}^\tau$ for the *n*-th band at momentum $\mathbf{k}$. The density of states (DOS) is then calculated by $\text{DOS}(\varepsilon_f) = \frac{1}{A}\sum_{nk\tau}\delta(\varepsilon_f - \varepsilon_{nk}^\tau)$, where $A$ is the system area and $\varepsilon_f$ is the Fermi energy. As we consider hole doping, the filling factor $\nu$ accounts the missing electrons per moiré unit cell and is calculated by $\nu = -A_0 \int_{\varepsilon_f}^{+\infty} \text{DOS}(\varepsilon)\, d\varepsilon$, where $A_0$ is the area of moiré unit cell. In our numerical calculation, we discretize the moiré Brillouin zone into a dense mesh of 270000 momentum points and approximate the delta function by $\delta(\varepsilon) \approx \frac{1}{\eta\sqrt{2\pi}} e^{-\frac{\varepsilon^2}{2\eta^2}}$ with the broadening parameter $\eta$ to be 5 times the maximum energy level spacing in the topmost moiré band on the discretized mesh.

**Exact Diagonalization Calculation**

In the ED calculation, we apply a particle-hole transformation to the moiré Hamiltonian in Eq. (1) and project the many-body Hamiltonian onto the first moiré band in $+K$ valley, assuming spontaneous full valley (spin) polarization. This band is topologically nontrivial and has a Chern number of 1. We take the gate-screened Coulomb potential as the interaction, which is $V(\mathbf{q}) = \frac{2\pi e^2}{\epsilon |\mathbf{q}|} \tanh(|\mathbf{q}|d)$ in momentum space. In the numerical calculation, we use a phenomenological dielectric constant $\epsilon = 10$ and gate-to-sample distance $d = 10$ nm. We find that the ED spectrum is not sensitive to the variation in $d$. The bandwidth of the first moiré band is 33 meV at $\theta = 4.54°$, while the characteristic interaction strength $\frac{e^2}{\epsilon a_M} \approx 32$ meV. These two energy scales are comparable, placing the system in the moderate correlated regime at $\theta = 4.54°$.

The states obtained in the ED spectrum are momentum eigenstates that follow the lattice translational symmetry. To construct a charge-density wave state that explicitly breaks this symmetry, we project an attractive impurity potential $\Delta(\mathbf{w}) = -\sum_{i=1}^{N} \delta(\mathbf{r}_i - \mathbf{w})$ onto the four-fold quasi-degenerate ground state manifold $\{|\Psi_j\rangle\}$. We choose the lowest-eigenvalue state of the projected $\Delta(\mathbf{w})$ with $\mathbf{w} = \mathbf{0}$, denoted as $|\Psi_{\text{QAHC}}\rangle$, to illustrate the charge order in the QAHC. Here $|\Psi_{\text{QAHC}}\rangle$ is a linear superposition of the four-fold quasi-degenerate ground states. The hole density shown in Fig. 3(i) is calculated with respect to $|\Psi_{\text{QAHC}}\rangle$.

**Acknowledgements**
X.L acknowledges support from National Key R&D Program (Grant Nos. 2024YFA1409002 and 2022YFA1403500), the National Natural Science Foundation of China (Grant Nos. 12521006 and 12274006) and Beijing National Laboratory for Condensed Matter Physics (Grant No. 2025BNLCMPKF001). M.W is supported from the National Natural Science Foundation of China (Grant No. 12404044). F.W is supported by National Key R&D Program (Grant Nos. 2022YFA1402400 and 2021YFA1401300) and the National Natural Science Foundation of China (Grant Nos. 12274333 and 12404084). S.L. acknowledges support from the Quantum Science and Technology – National Science and Technology Major Project (Grant No. 2023ZD0300500) and the Hong Kong RGC (Grant No. 26308524). This work is also supported by Peking Nanofab.

**Author contributions**
X.L. supervised the project; M.W. and L.L. fabricated the devices and performed the transport measurements with the help from Z.Z., Z.H., Q.Y.; M.W., L.L., Y.O., W.Q., F.W., and X.L. analyzed the data; Y.O., W.Q. and F.W. performed the theoretical modeling; K.W., T.T. and N.W contributed hexagonal boron nitride materials; Y.J and S.L. grew the $MoTe_2$ crystals. M.W., L.L, Y.O., W.Q., S.L., F.W and X.L. wrote the manuscript, with the input from all others.

**Competing interests**
The authors declare no competing interests.

# Figures

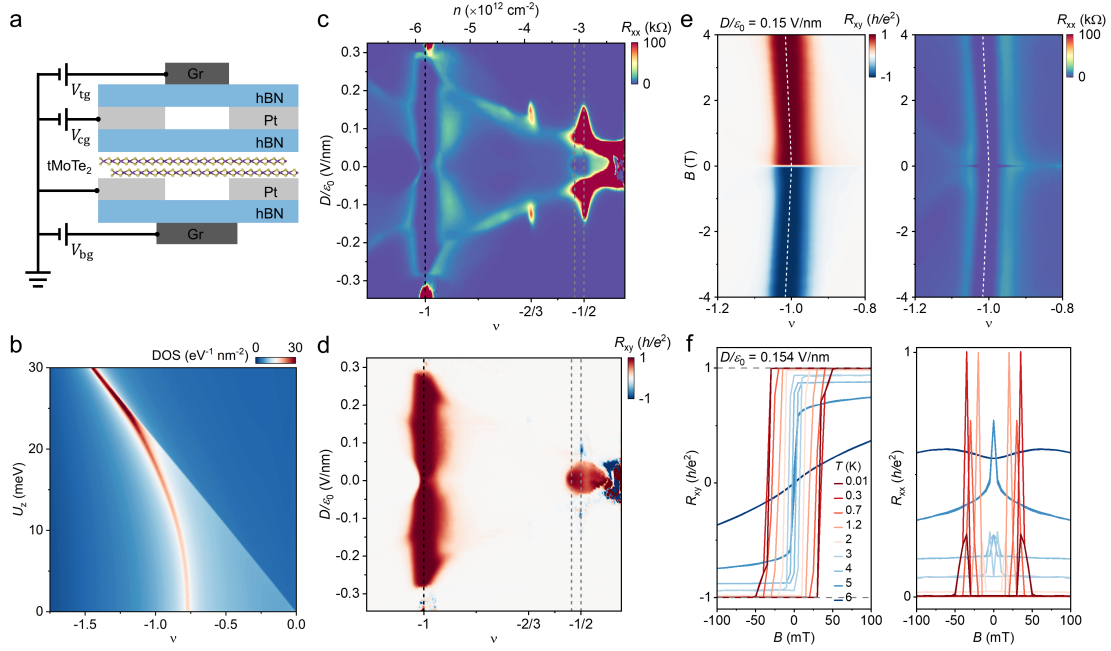

**Figure 1 | Phase diagram of moiré MoTe$_2$. a**, Schematic of the dual-gated tMoTe$_2$ device with Pt contact electrodes (light grey). Both top and bottom gates ($V_{tg}$ and $V_{bg}$) are made of few-layer graphite electrodes (dark grey) and hBN dielectrics. A local Pt contact gate ($V_{cg}$, light grey) is used to reduce the contact resistance for low-temperature measurements. **b**, Calculated single-particle DOS for 4.54° tMoTe$_2$, showing a continuous shift of vHS with $U_z$. **c**, **d**, Electric-field- and filling-factor-dependent $R_{xx}$ and $R_{xy}$ at $T = 10$ mK. The corresponding carrier density ($n$) is shown on the top axis. $R_{xx}$ and $R_{xy}$ are symmetrized and antisymmetrized, respectively, using the data taken at $B = \pm 10$ mT. Black and grey dashed lines mark $\nu = -1, -0.53$, and $-1/2$. **e**, Landau-fan diagram of $R_{xy}$ (left) and $R_{xx}$ (right) under $D/\varepsilon_0 = 0.15$ V/nm at $T = 10$ mK. The white dashed lines correspond to an IQAH state with Chern number $|C| = 1$ at $\nu = -1$. **f**, Magnetic-field-dependent $R_{xy}$ (left) and $R_{xx}$ (right) measured at different temperatures under $D/\varepsilon_0 = 0.154$ V/nm and $\nu = -1$. The quantized $R_{xy}$ and concomitant minimum in $R_{xx}$ persist up to 2 K, while the hysteresis disappears at 6 K. The grey dashed lines are to guide the eye, indicating the Hall resistance quantized at $|h/e^2|$.

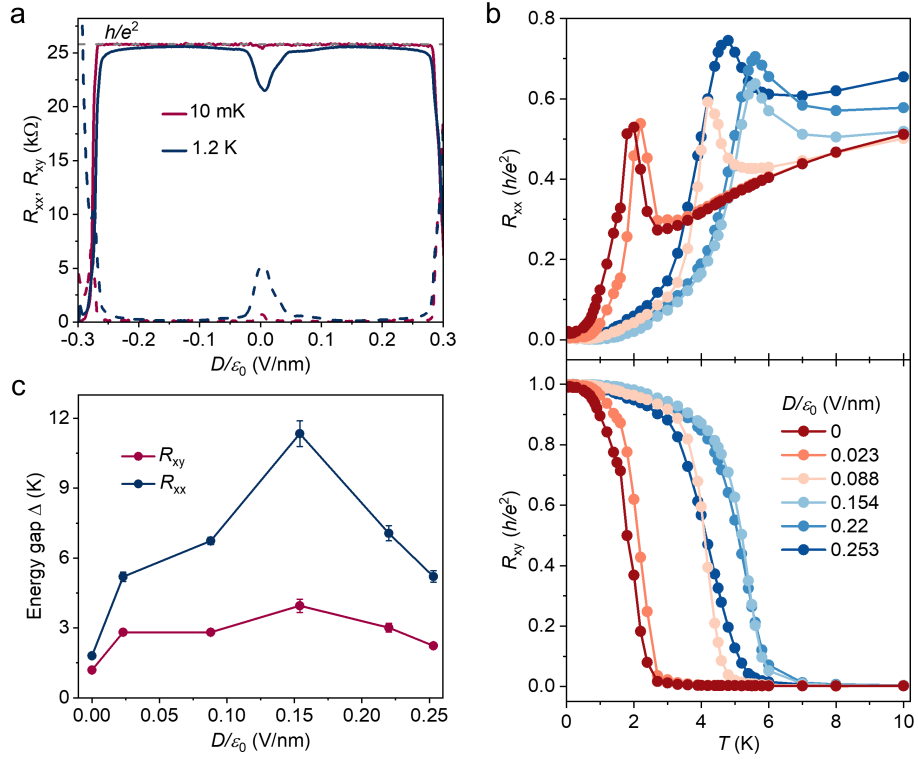

**Figure 2 | Electric-field-tuned IQAH effect at** $v = -1$. **a**, Symmetrized $R_{xx}$ (dashed lines) and antisymmetrized $R_{xy}$ (solid lines) as a function of $D/\varepsilon_0$ at $B = \pm 10$ mT and $T = 10$ mK (red) and 1.2 K (blue). **b**, Temperature dependence of $R_{xx}$ (top) and $R_{xy}$ (bottom) at selected electric fields under zero magnetic field. **c**, Energy gap as a function of electric field extracted from the data in panel **b** by using the thermal activation model (Extended Data Fig. 5). The error bars are uncertainties from the linear fits.

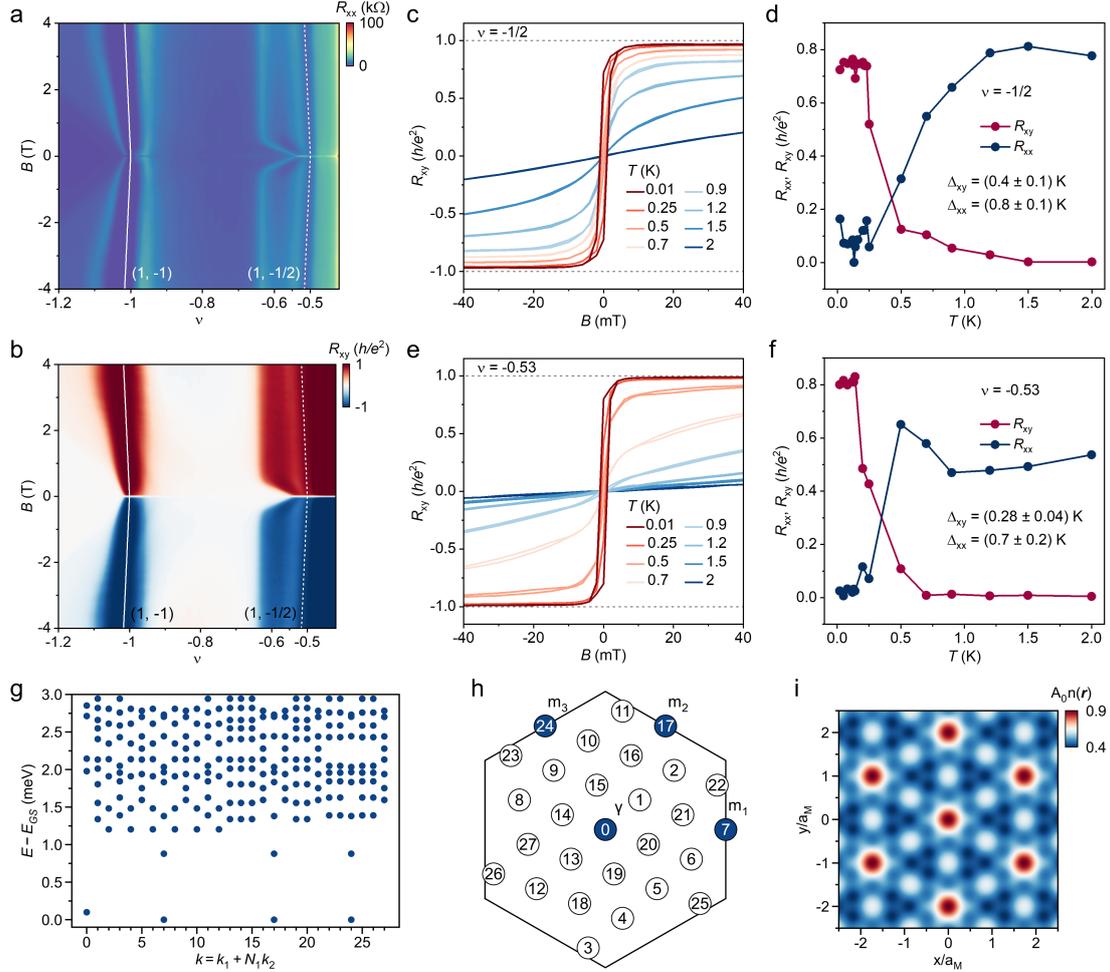

**Figure 3 | Topological states at $v = -1/2$ and $v = -0.53$. a, b,** $R_{xx}$ and $R_{xy}$ as functions of magnetic field and filling factor measured at $D/\varepsilon_0 = 0$ and $T = 10$ mK. The filling factors of the $R_{xx}$ minimum (**a**) and $R_{xy}$ maximum (**b**) disperse linearly with magnetic field, yielding slopes corresponding to Chern number $|C| = 1$ at $v = -1$ (white solid lines) and $v = -1/2$ (white dashed lines). **c, e,** Magnetic-field-dependent $R_{xy}$ at selected temperatures with $D/\varepsilon_0 = 0$ for filling factors $v = -1/2$ (**c**) and $v = -0.53$ (**e**). Compared with $v = -1/2$ state, the near-quantized $R_{xy}$ for $v = -0.53$ is strongly suppressed with increasing temperature at $B = \pm 10$ mT. **d, f,** Temperature dependence of the magnitudes of $R_{xx}$ and $R_{xy}$ at zero magnetic and electric fields. The data for $R_{xy}$ and $R_{xx}$ are extracted from panels **c** and **e**, as well as Extended Data Fig. 6. The energy gap determined from an Arrhenius activation model is indicated. **g,** ED energy spectrum at $v = -1/2$ on a 28 unit-cell cluster. The energies relative to the lowest eigenenergy are shown. **h,** The momentum-space mesh in the moiré Brillouin zone used in the ED calculations. The four quasi-degenerate ground states are located at the $\gamma$ point and the three $m_{1,2,3}$ points. **i,** Hole density n(**r**) of the many-body ground state pinned by an attractive impurity potential. $A_0$ represents the area of one moiré unit cell. Pronounced peaks appear at one quarter of MM sites on the moiré superlattice, indicating crystalline order with 2 × 2 enlarged unit cells.

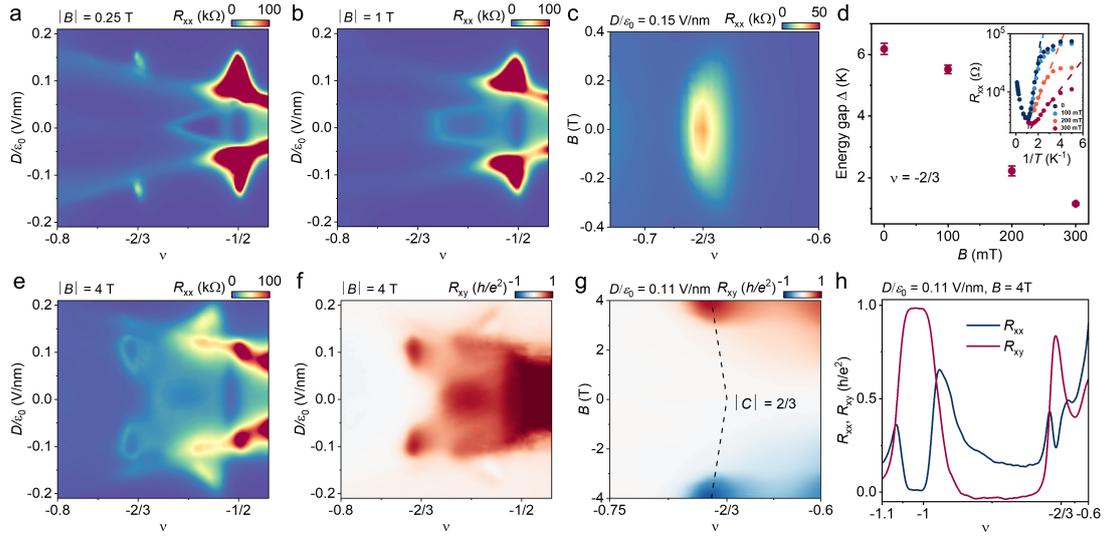

**Figure 4 | Magnetic-field-induced phase transition at $v = -2/3$. a, b,** $R_{xx}$ as a function of electric field and filling factor at $|B| = 0.25$ T (**a**) and 1 T (**b**). Both the states at $v = -2/3$ and at $v = -0.53$ exhibit pronounced magnetic field dependence. **c,** Evolution of insulator-metal transition at $v = -2/3$ measured at $D/\varepsilon_0 = 0.15$ V/nm. The correlated insulating state is observed for $|B| < 320$ mT. **d,** Magnetic-field-dependent energy gap extracted at $v = -2/3$ and $D/\varepsilon_0 = 0.15$ V/nm. The error bars represent the fit uncertainties from Arrhenius activation model (inset). **e, f,** $R_{xx}$ and $R_{xy}$ as a function of electric field and filling factor at $|B| = 4$ T. **g,** Landau-fan diagram of $R_{xy}$ measured under $D/\varepsilon_0 = 0.11$ V/nm. A Chern state emerges for $B > 3$ T, and disperses linearly with magnetic field. The slope is consistent with the dashed line, corresponding to a Chern number $|C| = 2/3$ determined by the Streda formula. **h,** Line-cuts of $R_{xx}$ and $R_{xy}$ at $B = 4$ T extracted from **g**. All data were collected at $T = 10$ mK, except for panel **d**.

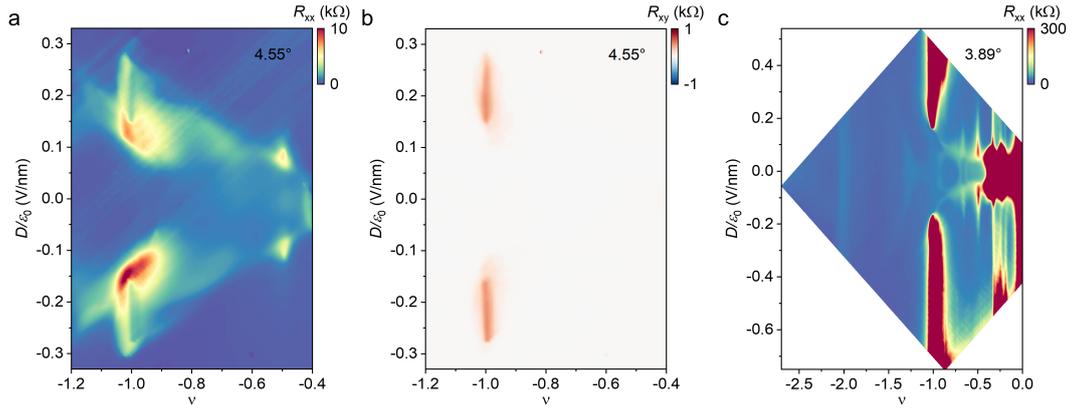

**Extended Data Figure 1 | Phase diagrams of device D2 and D3 at different twist angles. a**, **b**, $R_{xx}$ (**a**) and $R_{xy}$ (**b**) of device D2 (~ 4.55°) as a function of electric field and filling factor measured at $|B|$ = 0.2 T and $T$ = 10 mK. Signature of IQAH state is observed at $v = -1$. **c**, $R_{xx}$ of device D3 (~ 3.89°) as a function of electric field and filling factor measured at $B = 0$ and $T = 1.5$ K. This phase diagram is consistent with previous reports[13,14].

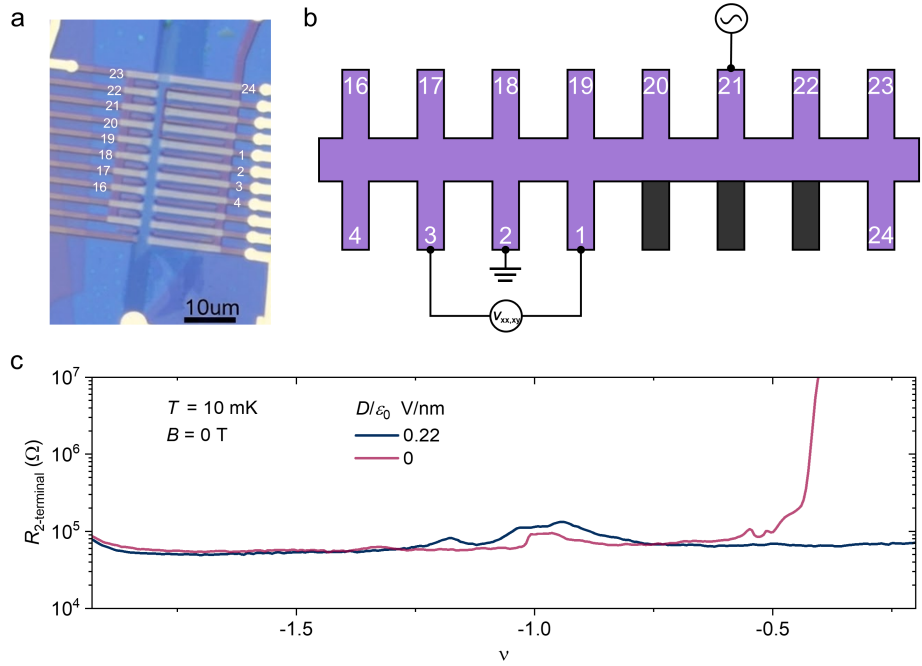

**Extended Data Figure 2 | Device architecture and low-temperature transport characteristics. a**, Optical image of the tMoTe$_2$ device D1. Scale bar, 10 μm. **b**, Schematic of the measurement configuration. A standard lock-in technique is employed, with an AC excitation applied between contacts 21 and 2. The voltage drop is probed between contacts 1 and 3. The three contacts shaded in black were etched. **c**, Two-terminal resistance $R_{2\text{-terminal}}$ as a function of the filling factor $v$, measured at a base temperature $T = 10$ mK and zero magnetic field. Data are collected at $D/\varepsilon_0 = 0.22$ V/nm (blue) and $D/\varepsilon_0 = 0$ (red). The resistance was measured between contacts 1 and 3 includes the contributions from the device and the contact resistances. Outside the insulating region, the total resistance saturates around 60 kΩ, consistent with a contact resistance of approximately 30 kΩ.

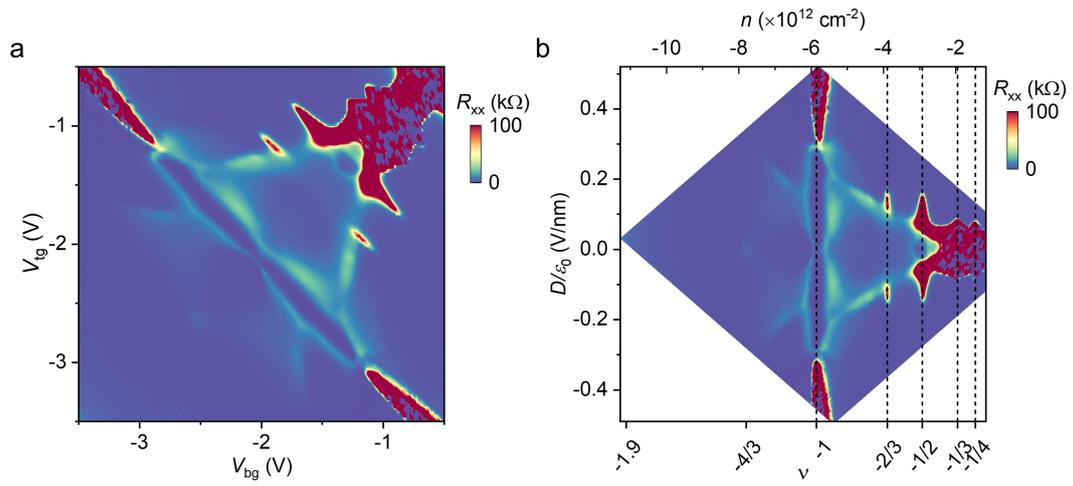

**Extended Data Figure 3 | Color maps of symmetrized longitudinal resistance $R_{xx}$ measured at $|B|$ = 10 mT and $T$ = 10 mK.** $R_{xx}$ is shown in $V_{bg}$–$V_{tg}$ (**a**) and $\nu$–$D/\varepsilon_0$ (**b**) parameter spaces. The black dashed lines delineate the correlated insulating states identified in 4.54° tMoTe$_2$.

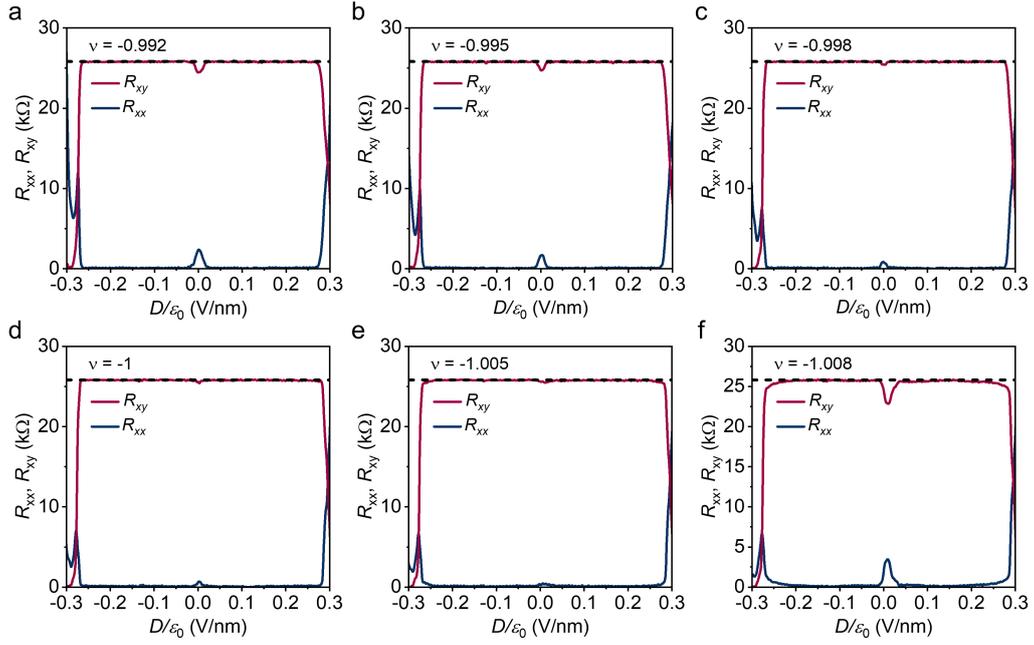

**Extended Data Figure 4 | IQAH state at different filling factors near $v = -1$.** Longitudinal resistance $R_{xx}$ (blue) and Hall resistance $R_{xy}$ (red) as functions of electric field measured at $T = 10$ mK for $v = -0.992$ (**a**), $-0.995$ (**b**), $-0.998$ (**c**), $-1$ (**d**), $-1.005$ (**e**) and $-1.008$ (**f**). As $v$ is tuned from $-1$, a deviation from quantization becomes evident at zero electric field. Black dashed lines are guides to the eye. The data are symmetrized ($R_{xx}$) and antisymmetrized ($R_{xy}$) at $|B| = 10$ mT.

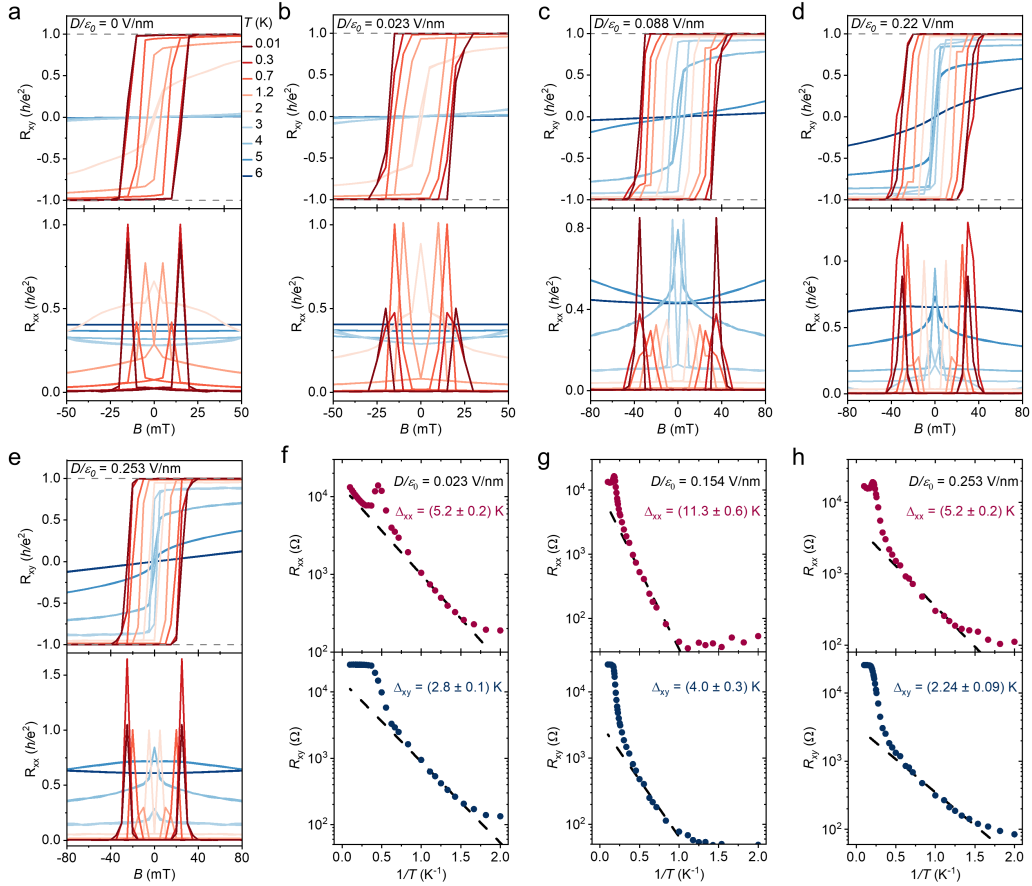

**Extended Data Figure 5 | Temperature dependence of the IQAH state at $v = -1$. a-e**, Evolution of the IQAH with temperature at selected electric fields $D/\varepsilon_0 = 0$ V/nm (**a**), 0.023 V/nm (**b**), 0.088 V/nm (**c**), 0.22 V/nm (**d**), and 0.253 V/nm (**e**). **f-h**, Energy gap extracted from Arrhenius analysis at $v = -1$. The resistance versus inverse of temperature with linear fits (black dashed lines) at different electric fields $D/\varepsilon_0 = 0.023$ V/nm (**f**), 0.154 V/nm (**g**), 0.253 V/nm (**h**). The corresponding energy gaps extracted from the fits are indicated.

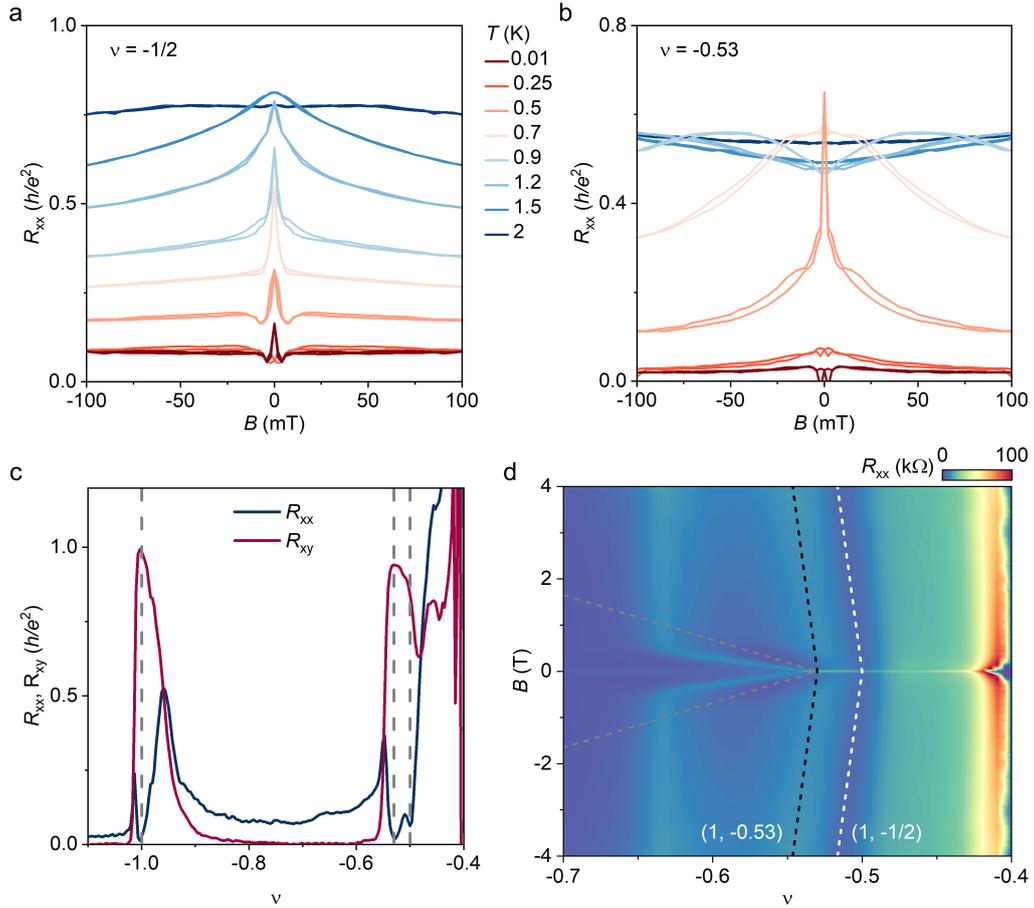

**Extended Data Figure 6 | Magnetic field dependence of $R_{xx}$ measured at $D/\varepsilon_0 = 0$.**
**a**, **b**, Evolution of $R_{xx}$ at $v = -1/2$ (**a**) and $-0.53$ (**b**) with temperature. **c**, Line-cuts of $R_{xx}$ and $R_{xy}$ at $T = 10$ mK taken from Fig. 1d. Grey dashed lines denote $v = -1$, $-0.53$ and $-1/2$. **d**, Landau fan diagram of $R_{xx}$ at $v = -1/2$ and $v = 0.53$ measured at $T = 10$ mK. White and black dashed lines indicate the linear dispersions in the $v-B$ maps for Chern number $|C| = 1$ at $v = -1/2$ and $v = -0.53$, respectively. A linear fit to $v = -0.53$ trajectory for $|B| < 1$ T (grey dashed line) yields a Chern number $|C| = 25$, which is inconsistent with the observed $R_{xy}$ in Fig. 3e.

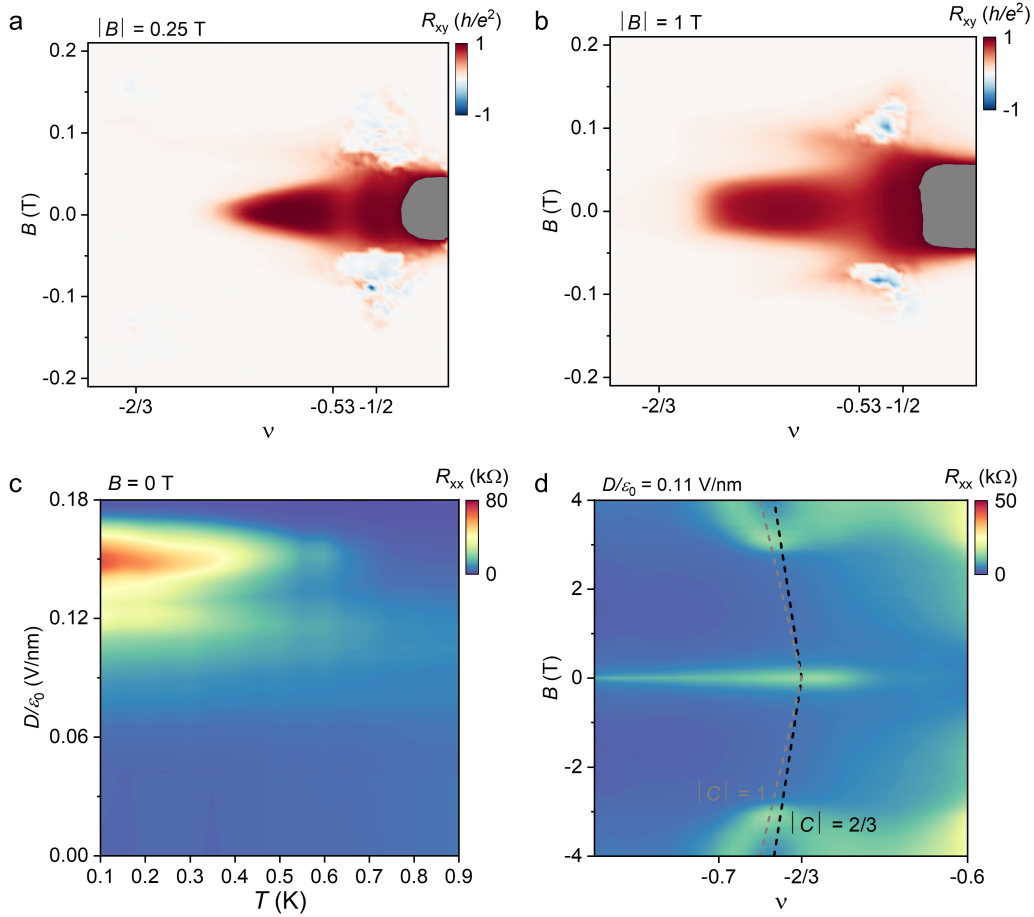

**Extended Data Figure 7 | $R_{xy}$ and $R_{xx}$ maps measured at $T$ = 10 mK. a, b,** $R_{xy}$ as a function of $v$ and $D/\varepsilon_0$ at $|B|$ = 0.25 T (**a**) and $|B|$ = 1 T (**b**). Compared with the $v = -1/2$, the state at $v = -0.53$ displays an anomalous magnetic field dependence. The grey shadowed regions indicate the Hall resistance exceed $|h/e^2|$. **c,** Longitudinal resistance $R_{xx}$ versus electric field $D/\varepsilon_0$ and temperature $T$ at $v = -2/3$. With increasing the temperature, the correlated insulating state near $D/\varepsilon_0$ = 0.15 V/nm is gradually suppressed. **d,** Longitudinal resistance $R_{xx}$ versus magnetic field $B$ and filling factor $v$ at $T$ = 10 mK. The dispersion of $R_{xx}$ minimum can be well captured by the Streda formula with $|C|$ = 2/3 (black dashed line) at $v = -2/3$, instead of $|C|$ = 1 (grey dashed line).